\documentclass[letterpaper]{article} % DO NOT CHANGE THIS
\usepackage{aaai2026}  % DO NOT CHANGE THIS
\usepackage{times}  % DO NOT CHANGE THIS
\usepackage{helvet}  % DO NOT CHANGE THIS
\usepackage{courier}  % DO NOT CHANGE THIS
\usepackage[hyphens]{url}  % DO NOT CHANGE THIS
\usepackage{graphicx} % DO NOT CHANGE THIS
\usepackage{natbib}  % DO NOT CHANGE THIS AND DO NOT ADD ANY OPTIONS TO IT
\usepackage{caption} % DO NOT CHANGE THIS AND DO NOT ADD ANY OPTIONS TO IT
\usepackage{algorithm}
\usepackage{algorithmic}

\usepackage{booktabs}
\usepackage{makecell}
\usepackage{multirow}
\usepackage{subfigure}
\usepackage{xcolor}
\usepackage{amsfonts}
\usepackage{amssymb}
\usepackage{amsmath}

\usepackage{newfloat}
\usepackage{listings}
\DeclareCaptionStyle{ruled}{labelfont=normalfont,labelsep=colon,strut=off} % DO NOT CHANGE THIS
\floatstyle{ruled}
\newfloat{listing}{tb}{lst}{}
\floatname{listing}{Listing}
\title{TableNet: A Large-Scale Table Dataset with LLM-Powered Autonomous Generation}
\author{
    Ruilin Zhang, Kai Yang\thanks{Corresponding author.}\\
    }
\affiliations{
    School of Computer Science and Technology, Tongji University\\
    4800 Cao'an Road, Shanghai, 201804, China\\

}

\usepackage{bibentry}
\begin{document}

\maketitle

\begin{abstract}
Table Structure Recognition (TSR) requires the logical reasoning ability of large language models (LLMs) to handle complex table layouts, but current datasets are limited in scale and quality, hindering effective use of this reasoning capacity. We thus present \textbf{TableNet dataset}, a new table structure recognition dataset collected and generated through multiple sources. Central to our approach is the first \textbf{LLM-powered autonomous table generation and recognition multi-agent system} that we developed. The generation part of our system integrates controllable visual, structural, and semantic parameters into the synthesis of table images. It facilitates the creation of a wide array of semantically coherent tables, adaptable to user-defined configurations along with annotations, thereby supporting large-scale and detailed dataset construction. This capability enables a comprehensive and nuanced table image annotation taxonomy, potentially advancing research in table-related domains. In contrast to traditional data collection methods, This approach facilitates the theoretically infinite, domain-agnostic, and style-flexible generation of table images, ensuring both efficiency and precision. The recognition part of our system is a diversity-based active learning paradigm that utilizes tables from multiple sources and selectively samples most informative data to finetune a model, achieving a competitive performance on TableNet test set while reducing training samples by a large margin compared with baselines, and a much higher performance on web-crawled real-world tables compared with models trained on predominant table datasets. To the best of our knowledge, this is the first work which employs the concept of active learning into the structure recognition of tables which is diverse in numbers of rows or columns, merged cells, cell contents, etc, which fits better for diversity-based active learning.
\end{abstract}

% Uncomment the following to link to your code, datasets, an extended version or similar.
% You must keep this block between (not within) the abstract and the main body of the paper.
\begin{links}
    % \link{Code}{https://github.com/Ha11oWe1t/TableNetExperimentCode}
    \link{Datasets}{https://huggingface.co/datasets/AnonymousUser123123/TableNet/tree/main}
%     \link{Extended version}{https://aaai.org/example/extended-version}
\end{links}

\section{Introduction}
\label{sec:intro}

Table Structure Recognition (TSR) aims to recover the logical structure of tables from images. Despite recent advances, it remains challenging due to the wide variability in table layouts, styles, and semantics. Real-world tables frequently include complex visual patterns—merged cells, missing borders, inconsistent alignments, or heterogeneous color schemes, which brings challenges to the logical reasoning ability to large language models (LLM) when parse the structure of tables. However, existing datasets are often limited in scale, diversity, and annotation quality, making it hard to fully exploit the logical reasoning ability of LLMs.

To address these limitations, we introduce the TableNet dataset and the first LLM-powered autonomous table generation and recognition multi-agent system. By integrating controllable parameters, the system is capable of generating a diverse range of tables with minimal human intervention. Unlike traditional dataset collection methods, our system leverages prior knowledge embedded in LLMs and is user-configurable, enabling the generation scalable, controllable, semantically coherent tables and corresponding annotations.

The broad style and domain coverage of TableNet motivates the use of active learning, which selects maximally informative samples for model training and could be used for data filtering in many fields like intrusion/anomaly detection\cite{zhu2019tripartite, yang2018active}, computer vision\cite{beluch2018power}, etc. This procedure can be described using a five-step algorithm shown in Algorithm \ref{alg:llm_active_learning}\cite{xia2025selection}, that is initialize, query, annotate, train and stop. The existing active learning method can be categorized into two main streams: diversity-based and uncertainty-based. Diversity-based methods\cite{gissin2019discriminative, sener2017active} aims to select a subset representative of overall distribution. Uncertainty-based methods\cite{beluch2018power, gal2017deep, yoo2019learning} prioritize samples that model cannot predict for sure. Diversity-based active learning is particularly suitable for heterogeneous table structures. By actively selecting training instances from multiple sources, our TSR model achieves competitive performance on the TableNet test set while requiring substantially fewer samples and significantly outperforms models trained on existing datasets when evaluated on unseen real-world tables.

Contributions of this paper can summarized as follows.

\begin{itemize}
\item We release TableNet, a large-scale table dataset composed of synthetic tables generated by a controllable LLM-based multi-agent system, real-world tables collected from the web, and augmented open source dataset, ensuring diversity in visual style, structure, and semantics. 
\item We develop the first LLM-based autonomous multi-agent system to generate table images with user-configurable properties, enabling synthesis of large-scale realistic dataset and serving as a fundamental support for TSR.
\item We employ a diversity-based active learning strategy to train the table structure recognition part of our system. Training on actively selected samples, our model achieves a competitive performance with much less training samples on its test set and a higher performance than models trained on other public datasets on web-crawled unseen real-world tables.
\end{itemize}

\begin{table*}
  \caption{\textbf{Comparison of TSR datasets.} For TableBank and SynthTabNet, we randomly sampled 1,000 images to assess color diversity and found that fewer than 10 were colored, indicating weak diversity. For TabRecSet and WTW, their diversity came from being captured by cameras in real world. \textbf{Annotation legend:} T: table bounding box (supports table detection); S: structural annotation (e.g., HTML tags, spanning); C: cell-level annotation (bounding box and content); H/X: final HTML/XML output or re-constructable from S and C; V: visual labels (e.g., simplicity, color, border style) provided in our dataset. \textbf{Definition of diversity:} A table image dataset is deemed as diverse if it supports at least two distinct and meaningful classification standards by which its images can be systematically categorized. These standards may include structural, visual attributes, allowing for multifaceted analysis and evaluation. For example, one classification standard could be based on structural complexity, such as whether a table contains spanning cells; another could be based on visual style, including attributes like border presence, background color, or watermark interference. A dataset that enables classification along both dimensions provides more comprehensive coverage for training and evaluating table structure recognition models.}
  \label{tab:dataset-comparison}
  \centering
  \begin{tabular}{llllll}
    \toprule
    Datasets        & Diversity      & Annotation & Domain                                    & Collection methodology                                                      & Tables \\
    \midrule
    TableNet (Ours) & True & S, C, H, V & \makecell[l]{Telecom\\(configurable)} & See Figure \ref{fig:collection-pipeline} & 445K \\
    PubTabNet       & False                & S, C, H    & Medicine                                  & rule-based collection & 510K   \\
    FinTabNet       & False                & T, S, C, H & Finance                                   & augmentation                                             & 113K   \\
    TableBank       & True & T, H       & General                                   & rule-based collection                    & 145K   \\
    SynthTabNet     & True & T, S, C, H & General                                   & augmentation                    & 600K   \\
    ICDAR 2019      & False                & T, X       & General                                   & institutional contribution                                                  & 3K     \\
    PubTables-1m    & False                & T, S, C, H & Medicine                                  & rule-based collection & 948K   \\
    TabRecSet       & True    & T, S, C, H & General                                   & manual labelling                                           & 38K    \\
    WTW             & True    & S, C       & General                                   & manual labelling                                           & 14K    \\
    SciTSR          & False                & T, C       & General                                   & rule-based collection                      & 15K     \\
    \bottomrule
  \end{tabular}
\end{table*}

\section{Related Works}
\label{sec:rw}

\textbf{Table Structure Recognition (TSR).} Tables are systematic arrangements of data organized into rows and columns, and their grid format provides clarity and efficiency. TSR aims to analyze the layout of table images and rebuild their cellular structure through representations such as markup languages \cite{zhong2020image, nassar2022tableformer}, spreadsheets \cite{koci2017table}, and graphs \cite{koci2019genetic, koci2018table, qasim2019rethinking, xue2019res2tim, chi2019complicated, liu2024grab}. Over the years, TSR approaches have evolved rapidly, progressed through three phases: (1) early approaches utilizing rule-based and heuristic methods, (2) deep learning-based methods, and (3) LLM-based methods.

Early approaches from the 1990s to 2010s relied heavily on visual cues such as cell boundaries and alignments. Pyreddy et al. \cite{pyreddy1997tintin} introduced the Character Alignment Graph using blank-space patterns, while Rus et al. \cite{rus1997customizing} proposed a similar White Space Density Graph. Several other rule-based or heuristic methods followed \cite{itonori1993table, kieninger1999t, shigarov2016configurable}. However, these techniques consistently struggled with complex table layouts, highlighting the need for more general solutions.

In the deep learning era, neural methods have dramatically surpassed early TSR approaches \cite{hashmi2021current}. Object detection–based methods \cite{zheng2021global, long2021parsing, schreiber2017deepdesrt, raja2020table, ajayi2024uncertainty, anand2023tc, gorishniy2021revisiting} identify structural components such as rows, columns, and captions. Markup-language models \cite{zhong2020image, nassar2022tableformer} treat TSR as image-to-markup translation using NLP-inspired architectures. Graph-based approaches \cite{koci2019genetic, koci2018table, qasim2019rethinking, xue2019res2tim, chi2019complicated, liu2024grab} represent tables as graphs to model cell-level relationships.

Recently, LLMs demonstrated exceptional ability in natural language related tasks. But LLM for TSR still remains an underexplored area and most work lies in table understanding field \cite{zheng2024multimodal, li2023table,zha2023tablegpt, su2024tablegpt2, pang2024uncovering, sui2024table}. Zhou et al. \cite{zhou2024enhancing} proposed Neighbor-Guided Toolchain Reasoner framework and significantly enhanced the recognition capabilities of vanilla vision LLMs. The availability of more datasets is likely to spur a rapid proliferation of LLM-based methods, fully leveraging the potential of large language models in handling tabular data.

\textbf{TSR Datasets.} Progress in Table Structure Recognition (TSR) has been largely driven by the release of benchmark datasets. The ICDAR series \cite{gobel2013icdar, gao2017icdar2017, gao2019icdar, kayal2021icdar} provides image-based datasets covering table detection, structure recognition, and end-to-end extraction, with table images ranging from PDFs to handwritten documents.

IBM Research has contributed several influential datasets. PubTabNet \cite{zhong2020image} focuses on TSR by converting table regions in PMCOA PDFs into HTML using the PubLayNet pipeline \cite{zhong2019publaynet}. FinTabNet \cite{zheng2021global} addresses the limitation of PubTabNet, which only supports TSR and lacks full-page context and overall table region bounding boxes. Furthermore, SynthTabNet \cite{nassar2022tableformer} further synthesizes PubTabNet, FinTabNet, and TableBank \cite{li2020tablebank} to overcome domain limitations and structural simplicity.

TableBank\cite{li2020tablebank} is another large-scale TSR dataset. It is created using weak supervision by automatically extracting tables from Word and LaTeX documents collected from the internet. This method leverages the structural information already embedded in these formats, allowing the dataset to be built at scale without extensive manual labeling. As a result, TableBank supports better generalization in TSR models, making them more effective in handling diverse and real-world table layouts.

As shown in Table~\ref{tab:dataset-comparison}, existing datasets still suffer from limited collection pipelines, domain constraints, and insufficient diversity. Models trained on one dataset often generalize poorly to others due to the diverse visual nature of tables \cite{li2020cross, somvanshi2024survey}. For example, a model trained on monochrome tables may perform poorly when applied to colored tables. Classic datasets such as UNLV \cite{cesarini2002trainable} and Marmot \cite{fang2012dataset}, along with recent ones like PubTables-1M \cite{yang2023large} and TabRecSet \cite{smock2022pubtables}, either lack the diversity \cite{zheng2021global, zhong2020image, nassar2022tableformer, li2020tablebank, cesarini2002trainable, fang2012dataset, yang2023large} or omit explicit table-type labels. \cite{long2021parsing, smock2022pubtables}. To address these limitations, we released TableNet, a large-scale synthesized TSR dataset, together with the first autonomous table generation multi-agent system to potentially facilitate TSR.

\textbf{Text-only LLM Image Synthesis.} Recent works have explored leveraging text-only LLMs to synthetic images for VLMs because of the lack of vision-language data \cite{cascante2022simvqa, johnson2016driving} and this approach has been applied to the generation of QA pairs of charts, plots. \cite{kahou2017figureqa, kafle2018dvqa} used small set of chart and fixed QA templates to generate chart VQA data. \cite{carbune2024chart, li2023scigraphqa} used text-only LLMs to generate annotations of questions or descriptions. Recent approaches like \cite{han2023chartllama, he2024distill, xia2024chartx, yang-etal-2025-scaling} generate charts or multiple kinds of image data, but they rely on LLMs to generate HTML code, making the overall procedure uncontrollable and more likely to generate error code. In our work, we designed a multi-agent system that explicitly decomposes the table generation into schema planning, layout construction and content filling, ensuring controllable table image synthesis and TSR annotating.

\section{Multi-Agent Table Generation System}
\label{sec:agent}

\begin{figure*}
  \centering
  \includegraphics[width=0.9\textwidth]{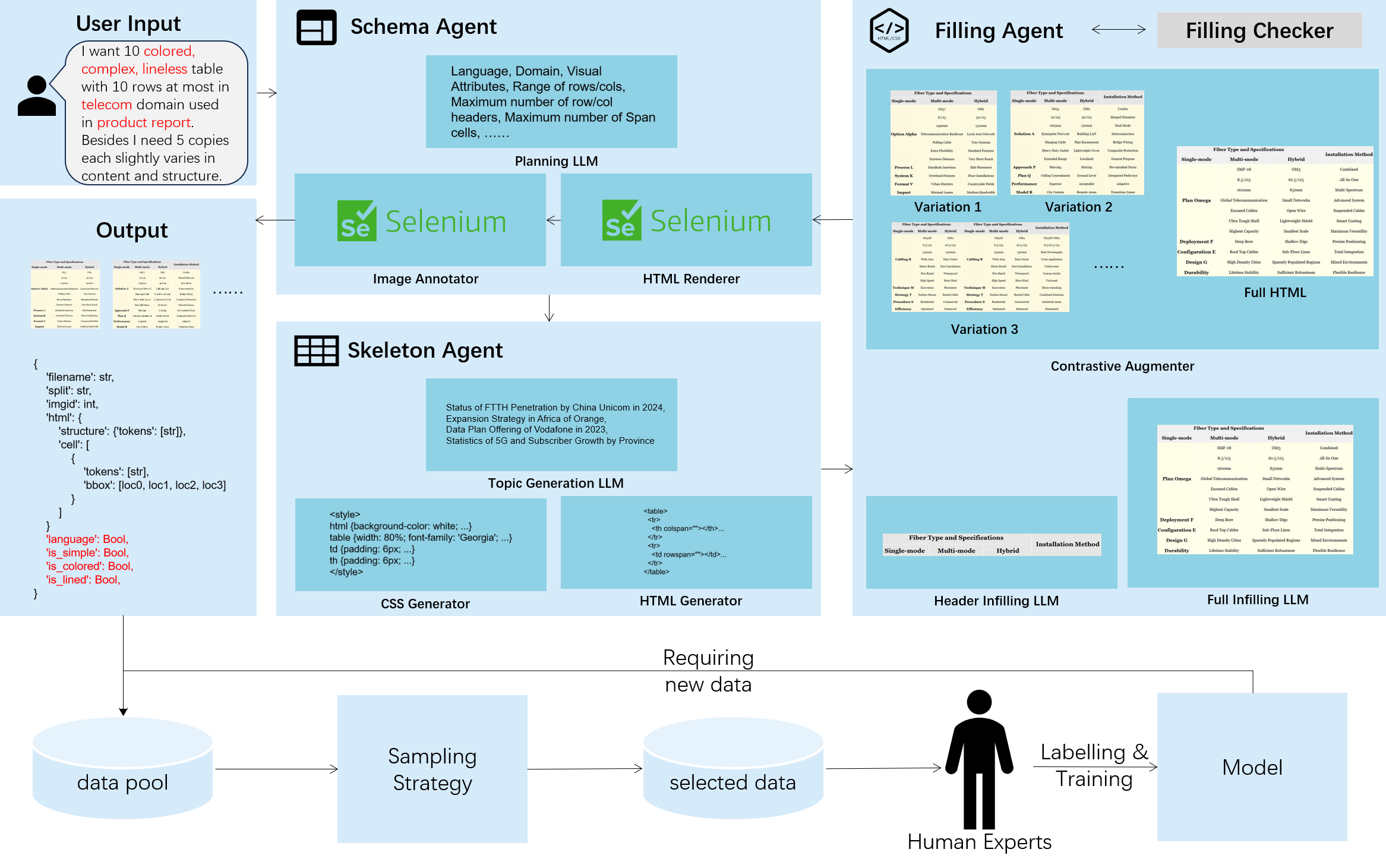}
  \caption{Workflow of our multi-agent system.}
  \label{fig:agent}
\end{figure*}

LLMs demonstrate competitive performance in various tasks\cite{jian2024tri, jian2025stable}, but passively predicting does not outperforms heuristic or simple agent tools on specific tasks like HTML structure generation. LLM-based multi-agent system is application that combine multiple LLMs with capabilities such as tool usage, planning, and memory. The LLM acts as the "brain" of the system interacting with users and execute key tasks. Planning helps decompose the request into manageable subtasks, which the system solves individually. Additionally, a reflection mechanism is integrated to refine execution. Tool usage enables the system to interact with the external and gather the necessary information to fulfill the user’s request. Since the workflow of our system is clearly illustrated in Figure~\ref{fig:agent}, therefore in this section, we describe how these capabilities are integrated into our table generation multi-agent, enabling configurable, domain-flexible, and semantically grounded table synthesis.

\subsection{Workflow and Core Concept Embodiment}
\label{sec:workflow}

\subsubsection{LLMs}

Our table generation system comprises three stage-specific agents coordinated by a core LLM. The core LLM interprets user requests—such as style, quantity, and domain—and orchestrates the workflow. A topic generation LLM produces table topics aligned with the specified or default telecommunication domain, ensuring semantic relevance. A header-infilling LLM and a body-infilling LLM then complete the HTML skeleton by inserting contents into the \texttt{\textless th\textgreater} and \texttt{\textless td\textgreater} tags. Our system is developed based on an open source codebase\footnote{https://github.com/WenmuZhou/TableGeneration/tree/main under MIT License.}.

\subsubsection{Planning}

The core LLM functions as a high-level planner that decomposes the task into structured subtasks with clear dependencies. Upon receiving a request, the Schema Agent determines table size and layout attributes, including row/column counts and spanning relations, and conditionally invokes a CSS generator for visual styling. It then generates a domain-relevant topic and constructs an initial HTML skeleton. Header and body infilling models are invoked sequentially after prerequisites are satisfied. Finally, the Filling Agent compares the filled and unfilled HTML to assess structural integrity and selectively regenerates the HTML if errors are detected. This multi-step, feedback-driven workflow ensures robust and controllable table synthesis.

\subsubsection{Tool Using}

To support multi-step execution, the system integrates several tool modules: (1) a CSS style generator; (2) an HTML tags generator specifying size and spanning; (3) a structure validator that builds a matrix representation and checks row equality; (4) a fallback HTML constructor for regenerating compliant tables; and (5) a Selenium tool for rendering table images and producing annotations.

\subsubsection{Memory}

The system incorporates a two-level memory mechanism. Outer memory retains the multi-turn dialog history between the core LLM and the user, ensuring continuity and refinement. Inner memory tracks previously generated table topics to avoid redundancy and promote semantic diversity. Together, these memories enhance interaction quality and generation diversity.

Within the Filling Agent, we adopt a contrastive learning–inspired augmentation strategy \cite{le2020contrastive} to enhance the diversity and robustness of TableNet. We define four transformations—copy, delete, swap, and alter—applied at row/column or block levels. Copy inserts duplicates; delete removes elements only when structural validity is preserved; swap exchanges rows/columns or blocks; alter adjusts row-level background colors. For each HTML structure, the body-infilling LLM generates multiple content variants (five without transformation and four with transformations), each assigned one operation. Row and column spanning regions are detected to prevent invalid modifications. This augmentation increases diversity and helps models learn fine-grained structural distinctions.

\begin{figure}
    \centering
    \includegraphics[width=0.23\textwidth]{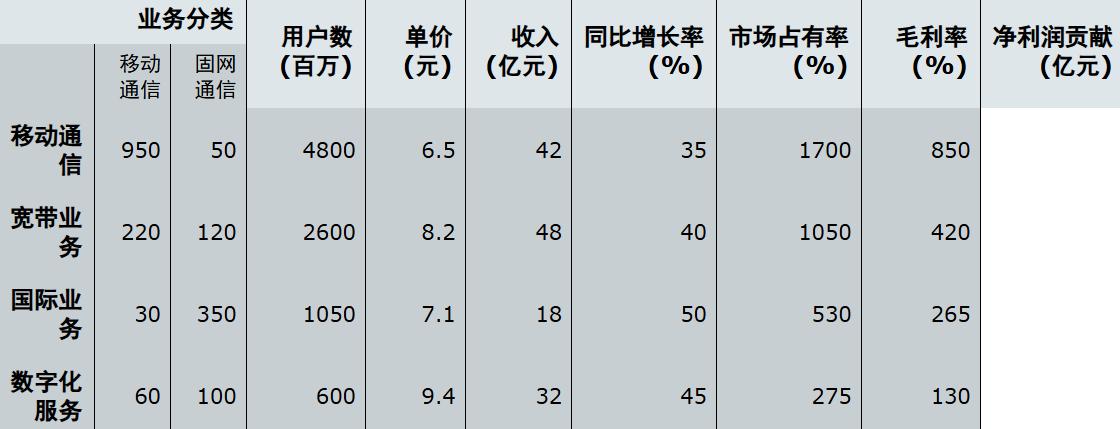}
    \hfill
    \includegraphics[width=0.23\textwidth]{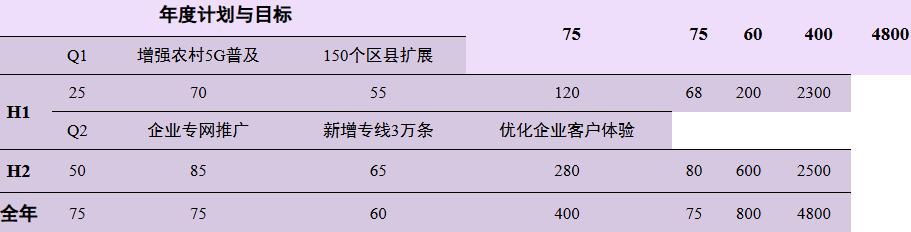}
    \caption{Invalid tables}
    \label{fig:invalid-table-example}
\end{figure}

To validate generated tables, we designed a mixed strategy filling checker to rank structure correctness, topic relevance, and semantic consistency. Structural correctness is evaluated using a heuristic method that checks validity (Figure \ref{fig:invalid-table-example}) and HTML tag misuse; topic relevance and semantic consistency are LLM-ranked. We verified that this filling checker can substitute for a well-trained human ranker.

After generation, we further use the data to finetune a TSR model under an active learning paradigm. This human-in-the-loop stage can be invoked at any time, since the data pool includes both labeled generated tables and unlabeled real-world tables, which are more valuable for TSR performance. We first apply a sampling strategy to select informative samples, then pass them to human experts for high-quality labeling and subsequent training.

\section{Dataset Collection and Composition}
\label{sec:colcomp}

\begin{figure*}
    \centering
    \includegraphics[width=0.7\textwidth]{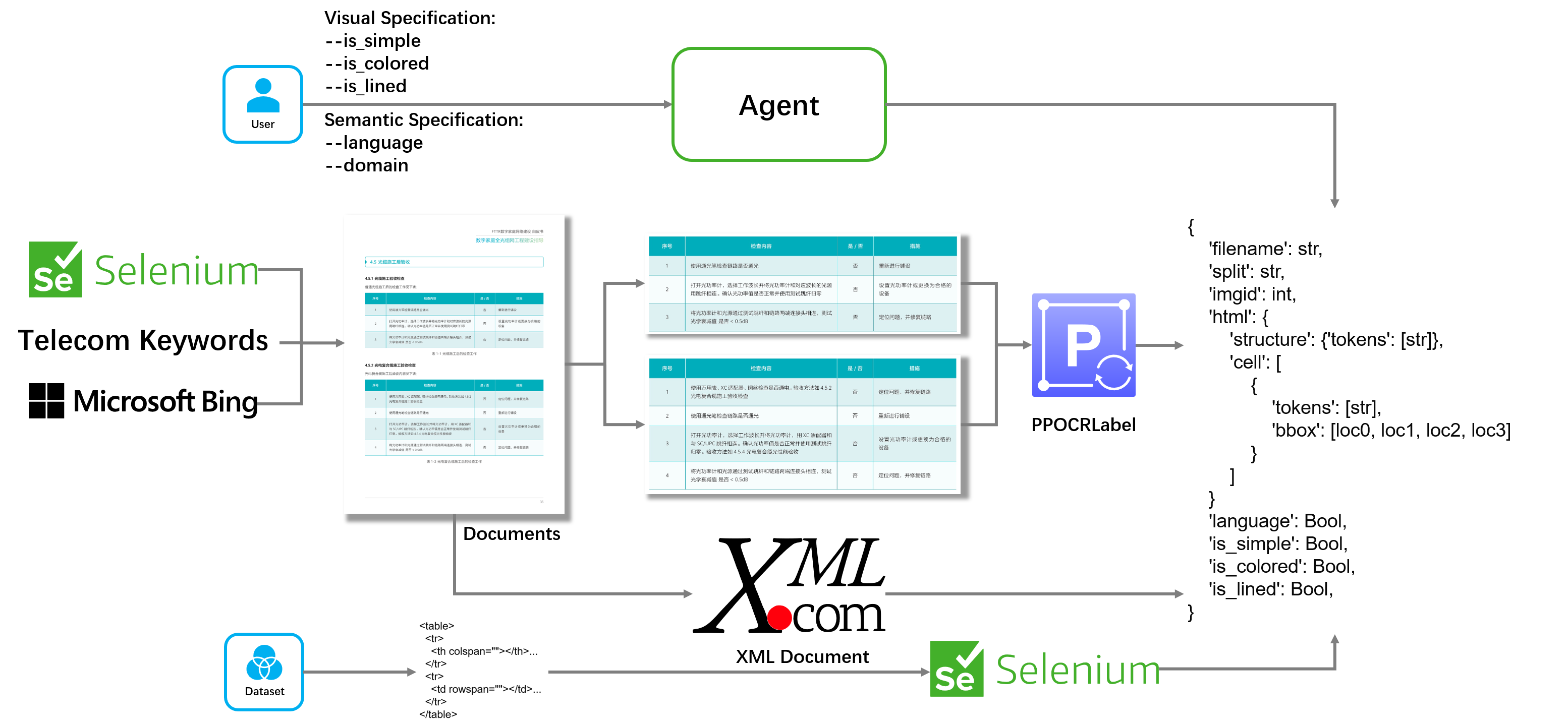}
    \caption{\textbf{Data collection pipeline.} From top to bottom: agent generating, web crawling and open source augmenting. \textbf{Annotation explanation:} is simple denotes whether the table contains cells spanning. is colored denotes whether the table includes any background color, colored borders, or non-black font colors. is lined specifies whether the cell borders are fully present; tables with only horizontal or vertical lines are considered not lined.}
    \label{fig:collection-pipeline}
\end{figure*}

\begin{figure*}
    \centering
    \subfigure[complex colored lined vertical table]{\includegraphics[width=0.24\textwidth]{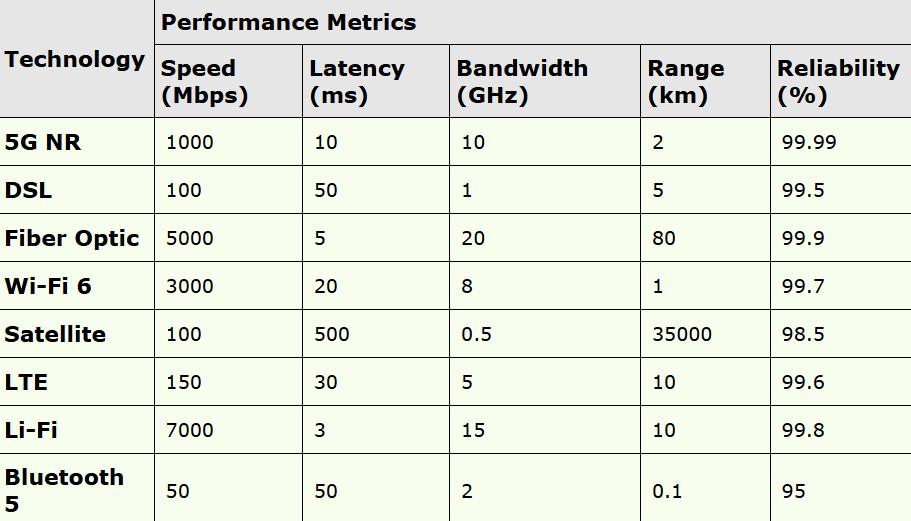}}
    \hfill
    \subfigure[complex zebra lineless matrix table]{\includegraphics[width=0.24\textwidth]{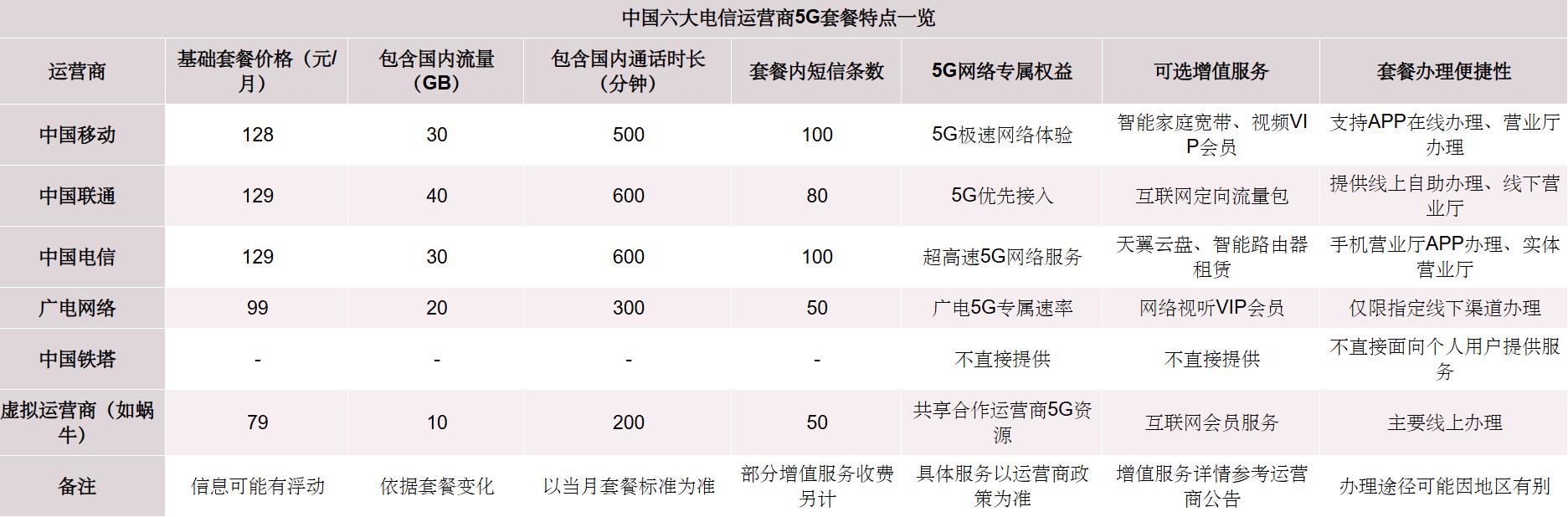}}
    \hfill
    \subfigure[simple colored lined vertical table]{\includegraphics[width=0.24\textwidth]{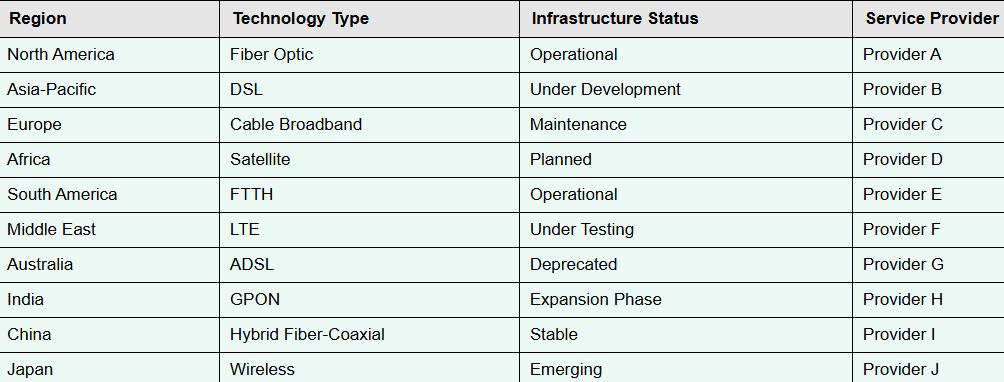}}
    \hfill
    \subfigure[simple zebra lined horizontal table]{\includegraphics[width=0.24\textwidth]{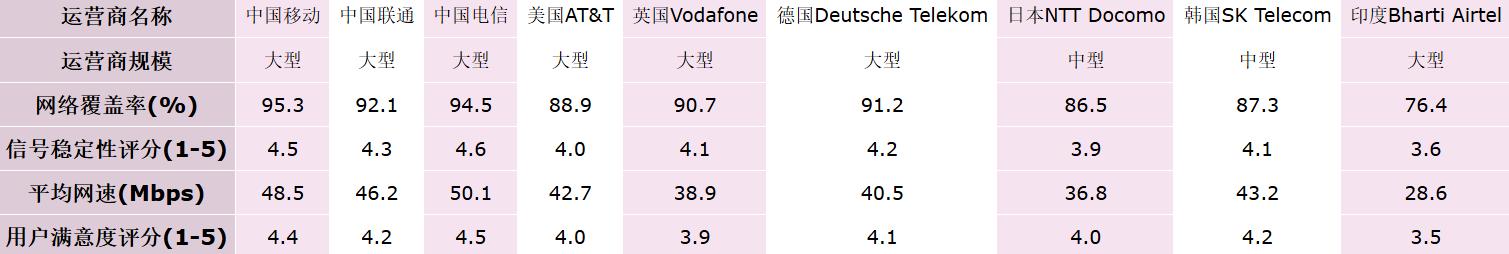}}
    \caption{examples in TableNet}
    \label{fig:dataset-examples}
\end{figure*}

Our dataset is constructed from three distinct sources: (1) agent-generated tables, (2) crawled tables from telecommunication-related PDF documents, and (3) crawled tables from Word documents, along with augmented open-source HTML table data. These three sources represent agent generation, manual labeling, and rule-based generation, respectively. The detailed dataset composition is available in Section \ref{sec:exp}.The pipelines and the final annotation format are illustrated in Figure \ref{fig:collection-pipeline}. We also present examples of tables in Figure \ref{fig:dataset-examples}.

\textbf{Agent-generating pipeline.} We employed an 8-way parallel generation strategy, where each process corresponds to a unique combination of three binary attributes (simple, colored, lined). In six days, we generated about 75K Chinese and 34K English tables, now expanded to 445K. To enhance realism, we target structural, semantic, and visual fidelity. Structurally, we summarize eight types of complex span-free tables based on orientation, multi-level headers, and body-level spans, all supported by our system. Semantically, we capture scenario-dependent complexity (e.g., tech comparisons vs. financial reports), and our filling checker improves authenticity by detecting header–body mismatches and hallucinations. Visually, we regulate border thickness, line style, font and background colors via CSS heuristics, and incorporate secondary factors such as resolution variation and watermarking noise.

\textbf{PDF and Word crawling \& labeling pipeline.} Given that China Telecom, China Mobile, China Unicom, and China Broadnet dominate the Chinese telecommunication industry, we used Selenium to construct queries combining their names with telecom-related keywords (e.g., FTTH, data plan;), restricting results to PDFs or Word files. For PDFs, due to the lack of reliable automatic extraction tools, we manually cropped and annotated 2.7K tables using PPOCRLabel over 30 days. For Word documents, we leveraged their zipped-XML structure to extract table markup, convert it to HTML, apply CSS via our generator, render images, and obtain annotations, yielding 600 tables despite the limited availability of Word files.

\textbf{Open-source dataset augmentation pipeline.} Since table understanding models typically take structured formats (e.g., HTML/Markdown) as input, we treated these structured inputs as outputs and rendered images and annotations using the same process as in the Word pipeline. Based on TABMWP \cite{lu2022dynamic}, we generated 1K augmented tables. However, augmentation alone is constrained by data availability, limited controllability, and restricted variability in domain, structure, and visual style.

\section{Experiments}
\label{sec:exp}

\subsection{Diversity and Filling Checker Analysis}

\textbf{Industry diversity groundness.} To validate the capability of our multi-agent system to generate tables of multiple domains, we selected 16 industries of 8 sectors from Global Industry Classification Standard (GICS) and calculated their Spearman and Pearson correlation and Kendall's tau between our filling checker and well-trained human rankers. Besides, we repeated the procedure three times and calculated the standard deviation to ensure the stability of the results. According to \cite{zhang2019bertscore}, our filling checker is capable of substituting well-trained human rankers and thus proves the semantic consistency and the multiple domain table generation ability. The detailed results are available as supplementary materials.

Across most sectors, the correlation metrics remain consistently high (mostly excels 0.8). Besides, we also evaluated average iterations for our system to correct a table and corresponding metrics after disturbing on structure, topic, and semantics in Table \ref{tab:checker-corr}. These indicate 1) strong semantic alignment within each language group and 2) filling checker is capable of ranking in replace of well-trained human rankers \cite{zhang2019bertscore}. This experiment provides a fine-grained view of the semantic cohesion within industrial table data and offers empirical grounding for further bilingual alignment or domain-adaptive table generation research.

\begin{figure}
    \centering
    \includegraphics[width=0.5\textwidth]{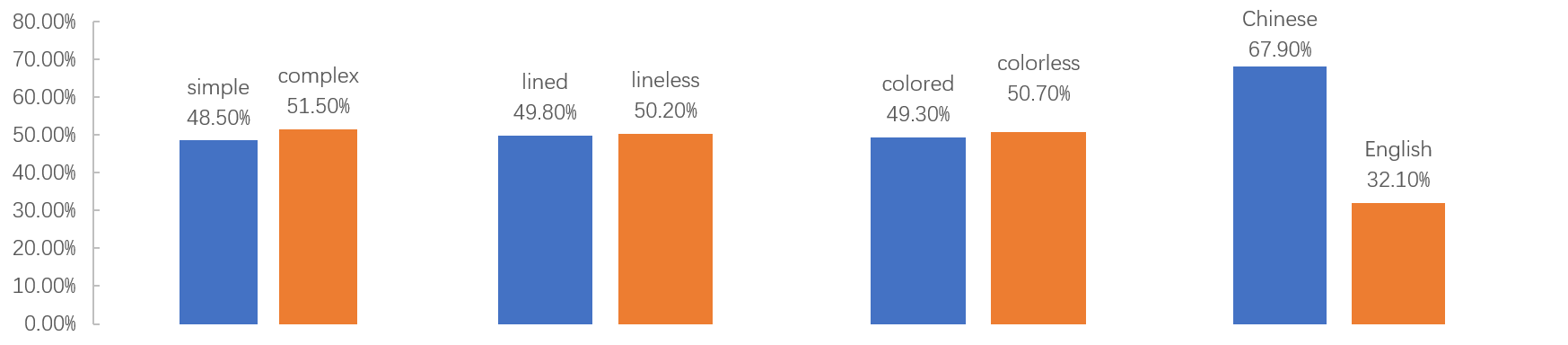}
    \caption{Generated data composition}
    \label{fig:dataset-composition}
\end{figure}

\textbf{Style diversity groundness.} By observeing our crawled real-world scenario tables, we described some key factor that determines the general veracity of a table. A distribution of configurable styles of TableNet is shown in Figure \ref{fig:dataset-composition}. As described in Section \ref{sec:agent}, our system is capable of generating multiple kinds of tables. For detailed table diversity, we evaluated detailed distribution of line style, structure complexity of our TableNet. Figure shown in supplementary materials indicates our TableNet covers most of given classifications where existing TSR datasets fail to achieve due to the limitations caused by collection methodology.

\begin{table}
  \caption{Correlation between human and filling checker ranks. Struct. and Sem. represent structure and semantics.}
  \label{tab:checker-corr}
  \centering
  \small
  \begin{tabular}{llllll}
    \toprule
    Complexity & Dim. & Avg Iters & Spear. & Pear. & Ken. \\
    \midrule
    \multirow{3}{*}{Simple}  & Struct. & 1.03 & 0.8203 & 0.8111 & 0.7928 \\
    & Topic & 1.07 & 0.8004 & 0.8085 & 0.6990 \\
    & Sem. & 1.03 & 0.9253 & 0.8914 & 0.8522 \\
    \multirow{3}{*}{Complex} & Struct. & 1.11 & 0.8826 & 0.8751 & 0.8832 \\
    & Topic & 1.08 & 0.7352 & 0.7408 & 0.6457 \\
    & Sem.  & 1.17 & 0.8753 & 0.8411 & 0.8019 \\
    \bottomrule
  \end{tabular}
\end{table}

\subsection{Experiment with LLMs}

We used TEDS (tree edit distance-based simlarity)\cite{nassar2022tableformer} as metric. To evaluate the effectiveness of our TableNet, we conducted a fine-tuning experiment using Qwen2-VL-2B on our TableNet Chinese train split, which contains about 48K tables. The full-parameter training process was conducted using 2 NVIDIA 4090 RTX GPUs with a batch size of 1 and a learning rate of 1e-4. The model was trained for 2 epochs, and the entire training process took approximately 8 hours. Training, testing split and corresponding subset for our training process will also be available. For baselines, we selected large-scale multimodal LLMs as baselines: Qwen2-VL-72B\cite{wang2024qwen2}, GPT\cite{achiam2023gpt}, Claude\cite{anthropic2024claude3} and Grok\cite{xai2024grok1} series. To avoid possible negative impact caused by inaccurate prompt design, we applied 1-shot and 5-shot in-context learning to Qwen2-VL-72B by integrating some training samples.

Table \ref{tab:llm-exp} summarizes the results. Models trained on TableNet show strong and stable performance across diverse table styles, whereas all baselines exhibit clear performance gaps between simple vs. complex structures, color vs. colorless layouts, and lined vs. lineless designs. This confirms that table diversity poses substantial generalization challenges and highlights the value of controllable generation and detailed annotations.Structural complexity—particularly row/column spans—remains the most difficult factor, as it requires precise spatial and hierarchical reasoning. Our fine-tuned Qwen2-VL-2B(FT) markedly narrows this gap and outperforms larger models on complex structures. Color variations also influence performance: while colored tables introduce richer context but higher variability, Qwen2-VL-2B(FT) remains robust (0.874 vs. 0.892), whereas zero-/few-shot large LLMs fluctuate more. Line presence further tests models’ ability to infer implicit structure; again, our model maintains consistent accuracy.Overall, even state-of-the-art LLMs struggle without targeted supervision, underscoring the necessity of structurally grounded training for TSR.

To further explained the effectiveness compared to other existing datasets, we used the crawled real-world tables as test set, and fine-tuned Qwen2-VL-2B on existing datasets using the same number of samples and identical training settings, and separately fine-tuned it on TableNet with crawled data excluded. We then evaluated models on unseen real-world tables spanning diverse structural, semantic, and visual styles. The model trained on TableNet achieved a TEDS of 0.7403, substantially outperforming the other-trained model, demonstrating the superior real-world generalizability brought by TableNet. The results are shown in Table \ref{tab:real-world}.

\begin{table}
    \caption{TSR performance on unseen real-world data of models trained on different existing datasets.}
    \label{tab:real-world}
    \centering
    \begin{tabular}{c|c}
        \toprule
        dataset & TEDS \\
        \midrule
       TableNet(Ours)  & \textbf{0.7403} \\
        PubTabNet & 0.5041 \\
        FinTabNet & 0.4495 \\
        SynthTabNet & 0.5242 \\
        TableBank & 0.5401 \\
        \bottomrule
    \end{tabular}
\end{table}

\begin{table*}
  \caption{TEDS of HTML on TableNet and subsets.}
  \label{tab:llm-exp}
  \centering
  \begin{tabular}{llllllll}
    \toprule
    \multirow{2}{*}{} & \multirow{2}{*}{} & \multicolumn{2}{c}{is simple} & \multicolumn{2}{c}{is colored} & \multicolumn{2}{c}{is lined}\\
    \cmidrule(r){3-4} \cmidrule(r){5-6} \cmidrule(r){7-8}
    Models & All & Simple & Complex & Colored & Colorless & Lined & Lineless \\
    \midrule
    \multicolumn{8}{c}{\textbf{Qwen}} \\
    \cmidrule(lr){1-8}
    Qwen2-VL-2B(FT) & \textbf{0.877} & 0.860 & \textbf{0.912} & \textbf{0.874} & \textbf{0.892} & \textbf{0.891} & \textbf{0.861} \\
    Qwen2-VL-72B & 0.721 & 0.822 & 0.604 & 0.726 & 0.713 & 0.711 & 0.729 \\
    Qwen2-VL-72B(1-shot) & 0.696 & 0.834 & 0.587 & 0.706 & 0.685 & 0.699 & 0.695 \\
    Qwen2-VL-72B(5-shot) & 0.688 & 0.791 & 0.576 & 0.695 & 0.682 & 0.689 & 0.687 \\
    \midrule
    \multicolumn{8}{c}{\textbf{GPT}} \\
    \cmidrule(lr){1-8}
    GPT-4.5-preview & 0.614 & 0.872 & 0.494 & 0.732 & 0.501 & 0.591 & 0.634 \\
    GPT-4o-all & 0.672 & 0.794 & 0.487 & 0.690 & 0.655 & 0.664 & 0.687\\
    \midrule
    \multicolumn{8}{c}{\textbf{Claude}} \\
    \cmidrule(lr){1-8}
    Claude-sonnet-4 & 0.620 & 0.904 & 0.517 & 0.687 & 0.533 & 0.621 & 0.618 \\
    Claude-sonnet-4-thinking & 0.706 & 0.917 & 0.561 & 0.693 & 0.719 & 0.749 & 0.648 \\
    Claude-opus-4 & 0.691 & \textbf{0.921} & 0.523 & 0.704 & 0.675 & 0.722 & 0.661 \\
    \midrule
    \multicolumn{8}{c}{\textbf{Grok}} \\
    \cmidrule(lr){1-8}
    Grok-4 & 0.697 & 0.892 & 0.538 & 0.731 & 0.642 & 0.678 & 0.707 \\
    \bottomrule
  \end{tabular}
\end{table*}

\subsection{Ablation Study}

LLM-driven data generation often suffers from structural instability, prompting us to evaluate LLMs’ end-to-end table synthesis capability using structure-only TEDS \cite{zhong2020image}. For simple tables, we instructed the LLM to produce HTML with specified row/column counts; for complex tables, we supplied explicit row- and column-span matrices that are difficult to express in natural language. As shown in Table \ref{tab:structure-only}, LLMs frequently fail to reproduce accurate structures—even in simple cases—while rule-based HTML generation reliably yields correct layouts and avoids structural infidelity through controlled prompting. These results demonstrate that, compared with directly generating tables via LLMs, an agent that invokes LLMs only at appropriate stages provides significantly more stable and accurate structure synthesis.

\begin{table}
  \caption{Table Fidelity using Structure-only TEDS of HTML generated by LLM through directly prompting. Small texts after ones means TEDS structure infidelity introduced by LLM infilling, which is lowest.}
  \label{tab:structure-only}
  \centering
  \begin{tabular}{lllll}
    \toprule
    \multirow{2}{*}{} & \multicolumn{2}{c}{Simple} & \multicolumn{2}{c}{Complex} \\
    \cmidrule(r){2-3} \cmidrule(r){4-5}
    Methods & with & without & with & without \\
    \midrule
    Qwen2.5 & 0.954 & 0.978 & 0.782 & 0.835 \\
    DeepSeek-R1 & 0.938 & 0.974 & 0.940 & 0.901 \\
    GPT-4o & 0.980 & 0.978 & 0.931 &  0.928 \\
    Agent Tool & 1 \small-0.004 & 1 \small-0.003 & 1 \small-0.012 & 1 \small-0.009 \\
    \bottomrule
  \end{tabular}
\end{table}

\begin{figure}
    \centering
    \includegraphics[width=0.45\textwidth]{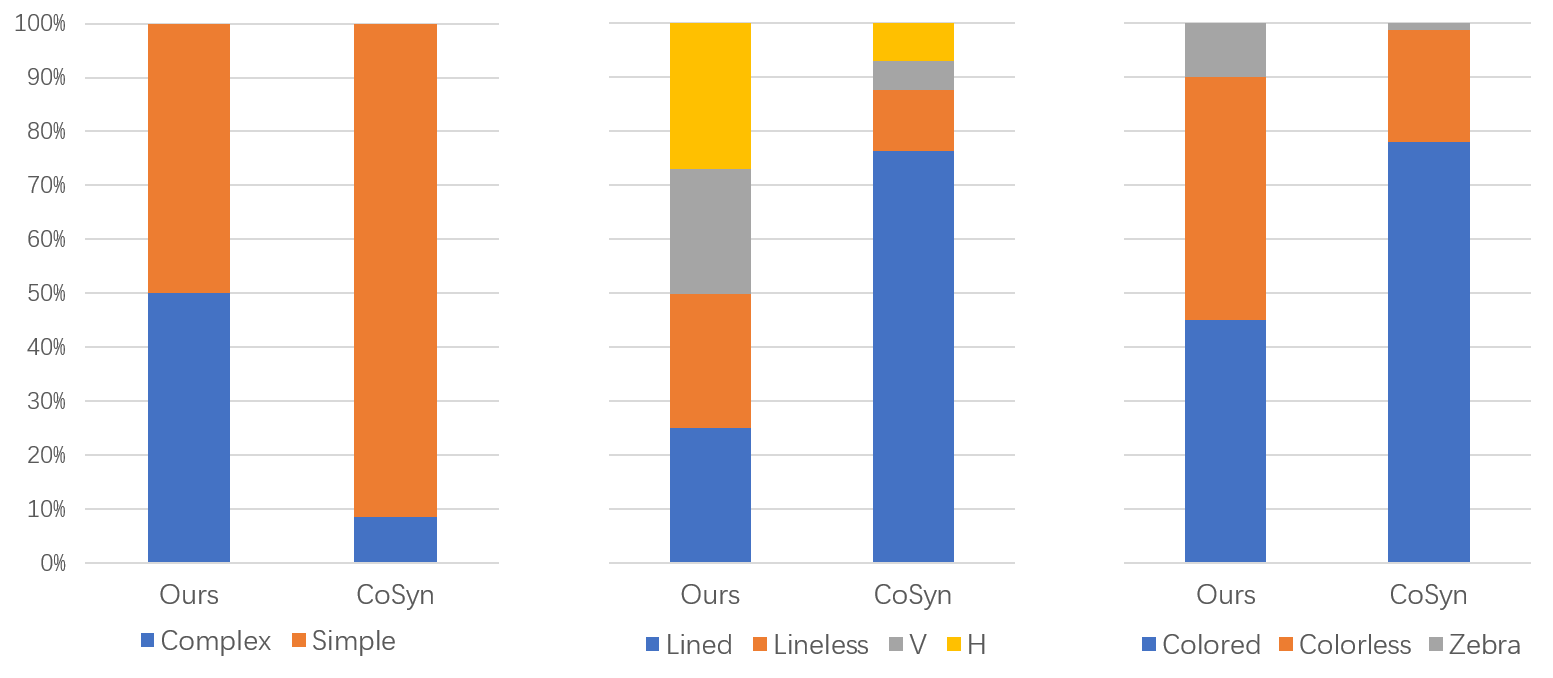}
    \caption{Compared with existing CoSyn pipeline, which is an LLM end-to-end method. In the middle subfigure, H represents horizontally lined and V represents vertically lined.}
    \label{fig:ab}
\end{figure}

To assess whether agent-based design improves dataset quality, we evaluated the recent CoSyn pipeline for table-image and QA generation \cite{yang-etal-2025-scaling}. Using it to generate 1,000 tables, only 85 contained complex structures such as merged rows/columns; most followed a simple stacked \verb|<tr><td>| format. We also evaluate the color and line style diversity, results are shown in Figure \ref{fig:ab}.

After strengthening its HTML prompt to enforce structural complexity, the enhanced baseline achieved an average structural score of 4.31 (out of 5) under our checker. In contrast, our agent-based pipeline reached 4.79 on average and produced structurally correct tables with only 1.11 fallback iterations per sample.

\subsection{Experiment on Active Learning TSR}

The active learning procedure is detailed in the following. We use an multi-agent system to generate new data instances and corresponding labels, not just selecting from the unlabeled set $\mathcal{U}$. The LLM serves as a powerful tool throughout the AL loop, starting by annotating an initial dataset ($\mathcal{L}_{\text{init}}$) to address the cold-start problem. In the Query step, we selects informative instances from $\mathcal{U}$. These instances are then annotated in the Annotation step, either automatically or manually. The newly labeled data is added to the labeled set $\mathcal{L}$, and the target model $f_\theta$ is updated. The process iterates until the given annotation budget is met. This method overcomes challenges in traditional AL by using LLMs for data generation and automatic agent tools for annotation, reducing reliance on human annotators.

Based on Qwen2-VL-2B(FT) (hereafter the base model), we conducted an active-learning fine-tuning study using a diversity-based strategy because TableNet exhibits diversity in row/column counts, merged-cell locations, colors, etc. Training used 2 × NVIDIA RTX 4090 GPUs with LoRA fine-tuning for 1 epoch; TEDS served as the evaluation metric. We adopted CoreSet \cite{sener2017active}, a greedy k-center shown in Algorithm \ref{alg:kcenter} where we try to choose $b$ centers such that the largest distance between a data point and its nearest center is minimized, formally trying to solve shown below, where $s_{0}$ denotes the unlabeled dataset, $b$ is the annotation budget, and $s_{1}$ represents the set of samples selected for annotation.

\begin{equation}
\min_{s_{1}\,:\,|s_{1}| = b} \;
\max_{i} \;
\min_{j \in s_{1} \cup s_{0}} 
\Delta(x_i, x_j).
\end{equation}

We use it as a querying strategy shown in Algorithm \ref{alg:llm_active_learning}, and compared it to random sampling (RSB), hard-example mining (HE) \cite{liu2025hard, shrivastava2016training}, and perplexity-based uncertainty sampling (PPL). We extracted the patched vision embeddings $e \in \mathbb{R}^{patch \times dim}$ from the vision tower of base model and concatenated its max- and mean-pooled embedding into a vector $c \in \mathbb{R}^{2 \times dim}$ to perform the method.

\begin{algorithm}[t]
\caption{\textnormal{Active Learning}}
\label{alg:llm_active_learning}
\begin{algorithmic}[1]
\STATE \textnormal{\textbf{Input:}} \textnormal{Unlabeled dataset } $\mathcal{U}$, \textnormal{Model } $M$, \textnormal{Budget } $k$
\STATE \textnormal{\textbf{Output:}} \textnormal{Trained model } $f_{\theta}$, \textnormal{Labeled dataset } $\mathcal{L}$

\STATE $\mathcal{L}_{\text{init}}, \mathcal{U} \leftarrow \operatorname{Initialize}$
\STATE $f_\theta \leftarrow \operatorname{Train}(\mathcal{L}_{\text{init}})$

\WHILE{\textnormal{not} Stop$(k, f_\theta, M)$}
    \STATE $x \leftarrow \operatorname{Query}(f_\theta, \mathcal{U}, M)$
    \STATE $(x, y) \leftarrow \operatorname{Annotate}(x, M)$
    \STATE \textbf{if} $x \in \mathcal{U}$ \textbf{then} $\mathcal{U} \leftarrow \mathcal{U} \setminus \{x\}$
    \STATE $\mathcal{L} \leftarrow \mathcal{L} \cup \{(x, y)\}$
    \STATE $f_\theta \leftarrow \operatorname{Train}(\mathcal{L})$
\ENDWHILE

\STATE \textnormal{\textbf{return}} $f_\theta^\ast, \mathcal{L}$
\end{algorithmic}
\end{algorithm}

\begin{algorithm}[t]
\caption{\textnormal{k-Center-Greedy}}
\label{alg:kcenter}
\begin{algorithmic}[1]
\STATE \textnormal{\textbf{Input:}} \textnormal{Data points } $\{x_i\}_{i=1}^n$, \textnormal{initial set } $s_0$, \textnormal{budget } $b$
\STATE \textnormal{\textbf{Output:}} $s$

\STATE $s \leftarrow s_0$

\REPEAT
    \STATE $u \leftarrow \displaystyle \arg\max_{i \in [n] \setminus s} \; \min_{j \in s} \Delta(x_i, x_j)$
    \STATE $s \leftarrow s \cup \{u\}$
\UNTIL{$|s| = b + |s_0|$}

\STATE \textnormal{\textbf{return}} $s\setminus s_0$
\end{algorithmic}
\end{algorithm}

\begin{figure}
    \centering
    \includegraphics[width=0.4\textwidth]{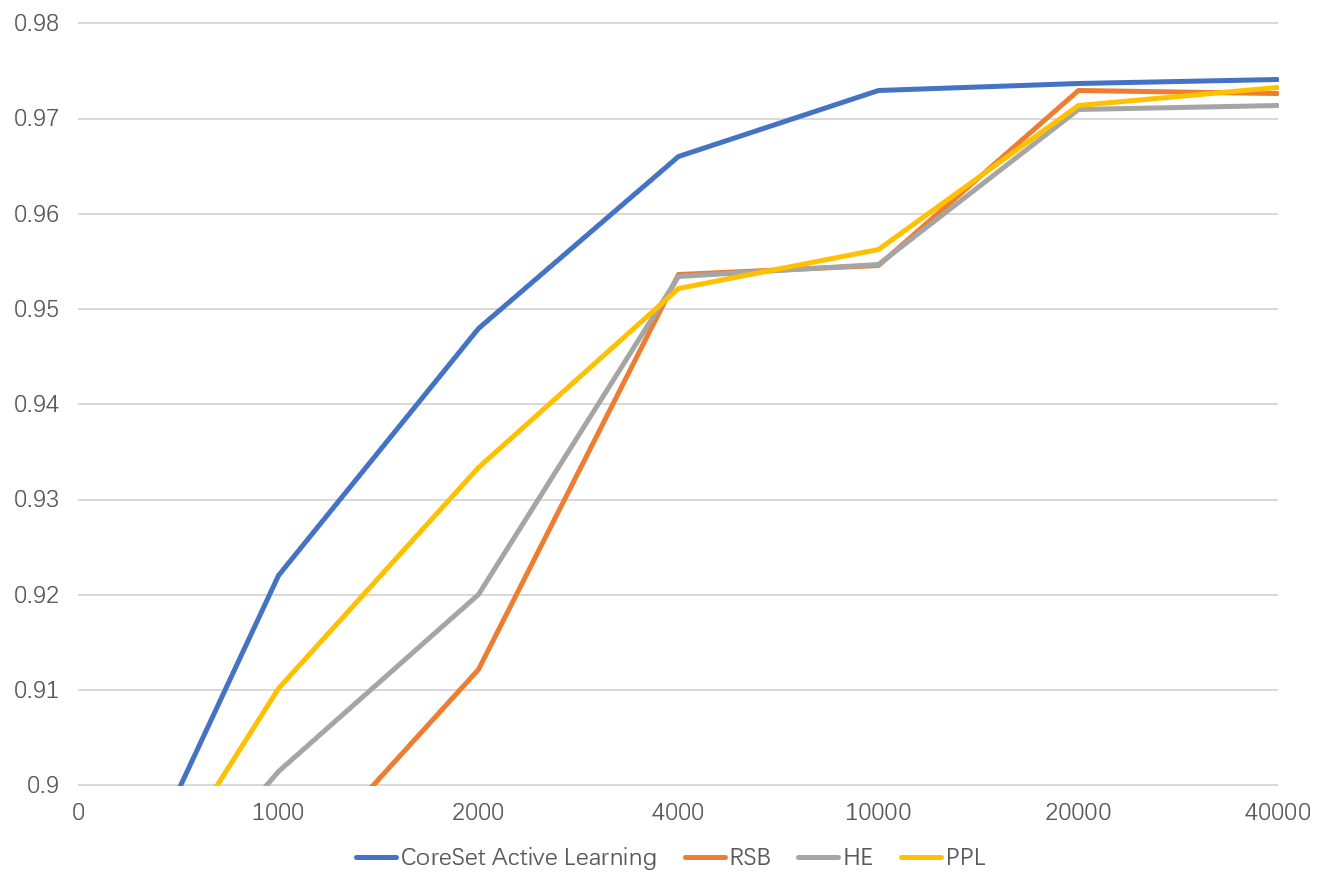}
    \caption{TSR Experiment results}
    \label{fig:TSR exp}
\end{figure}

As shown in Figure \ref{fig:TSR exp}, diversity-based CoreSet method demonstrates stably higher performance that baselines, including random sampling, perplexity-based uncertainty sampling, and hard examples mining. Besides, We can also observe that active learning method achieved a comparable performance compared to baselines while reducing the number of sample by at least $50\%$ in both low-data regime and standard-data regime by comparing the number of training samples of the same performance. For example, model finetuned on 10k actively selected samples achieved a TEDS performance around 0.973 while other baselines used more than 20k, even 40k training samples. 

\section{Limitations}
\label{sec:lim}

Although our system supports configurable parameters, it is still bounded by the pretraining distribution and reasoning capabilities of the underlying LLM. As a result, the diversity of generated content may not fully capture edge cases or highly domain-specific table formats. Furthermore, the quality of generated table images may vary due to model's knowledge of user specified domain and language, leading to irrelevant table contents or inconsistencies within semantic contents.

\section{Conclusion}
\label{sec:conclusion}

In this work, we introduce TableNet, a new dataset for TSR mainly sourced from autonomous generation. Complementing the dataset, we develop the first autonomous table generation and recognition multi-agent system that facilitates TSR and supports controllable table synthesis and automatic annotation through user-defined parameters. It enables scalable, flexible, and research-oriented data construction, and demonstrates high efficiency with the capacity to generate large volumes of structurally reliable tables. The scalability of our collection process, together with empirical results, highlights the significance of both TableNet and the system as TSR tools for TSR. We believe that integrating a powerful generation agent with TableNet will meaningfully push TSR.

\small
\bibliography{aaai2026}

\appendix
\section{Supplementary Experiment Results}
\subsection{Experiments on Multiple Industries}

To evaluate the semantic consistency of tables across different Global Industry Classification Standard (GICS) sectors and industries, we conducted a comprehensive correlation analysis between Chinese and English tables. Specifically, we computed the average pairwise semantic correlation between corresponding tables in the two languages under each industry category. The metrics include Spearman's rank correlation (Spear.), Pearson correlation (Pear.), and Kendall's Tau (Ken.), each reported as mean ± half standard deviation to reflect statistical stability. Results are shown in Table \ref{tab:domain-corr}

Each industry is evaluated separately, with two rows per industry: the first row corresponds to the semantic correlations for Chinese tables, and the second row to their English counterparts.

\begin{table*}
  \caption{Average semantic correlation adding or subtracting half standard deviation of multiple Global Industry Classification Standard (GICS) industries' tables. For each industry, the upper results line represents Chinese tables and the other line represents English tables.}
  \label{tab:domain-corr}
  \centering
\begin{tabular}{lllll}
\toprule
Sector                                  & Industry                                    & Spear. & Pear.  & Ken.   \\
\midrule
\multirow{4}{*}{Energy}                 & \multirow{2}{*}{Equipment \& Service}       & 0.8021 \small± 0.015 & 0.7954 \small± 0.007 & 0.7311 \small± 0.007 \\
                                        &                                             & 0.8204 \small± 0.016 & 0.8080 \small± 0.015 & 0.7551 \small± 0.010 \\
\cmidrule{2-5}
                                        & \multirow{2}{*}{Consumable Fuel}            & 0.8344 \small± 0.021 & 0.8047 \small± 0.010 & 0.7612 \small± 0.005       \\
                                        &                                             & 0.8121 \small± 0.002 & 0.7853 \small± 0.013 & 0.7224 \small± 0.018 \\
\cmidrule{1-5}
\multirow{4}{*}{Material}               & \multirow{2}{*}{Chemicals}                  & 0.8006 \small± 0.017 & 0.8105 \small± 0.015 & 0.7290 \small± 0.022 \\
                                        &                                             & 0.8788 \small± 0.027 & 0.8446 \small± 0.017 & 0.8054 \small± 0.020 \\
\cmidrule{2-5}
                                        & \multirow{2}{*}{Metals \& Mining}          & 0.8951 \small± 0.027 & 0.8848 \small± 0.028 & 0.8109 \small± 0.019 \\
                                        &                                             & 0.9091 \small± 0.006 & 0.9004 \small± 0.005 & 0.7923 \small± 0.016 \\
\cmidrule{1-5}
\multirow{4}{*}{Health Care}            & \multirow{2}{*}{Biotechnology}              & 0.8516 \small± 0.016 & 0.8476 \small± 0.015 & 0.7577 \small± 0.015 \\
                                        &                                             & 0.8607 \small± 0.016 & 0.8394 \small± 0.017 & 0.7509 \small± 0.004 \\
\cmidrule{2-5}
                                        & \multirow{2}{*}{Pharmaceuticals}            & 0.8585 \small± 0.007 & 0.8347 \small± 0.010 & 0.7544 \small± 0.008 \\
                                        &                                             & 0.8433 \small± 0.015 & 0.8172 \small± 0.011 & 0.7712 \small± 0.017 \\
\cmidrule{1-5}
\multirow{4}{*}{Financials}             & \multirow{2}{*}{Banks}                      & 0.8821 \small± 0.018 & 0.8543 \small± 0.019 & 0.7654 \small± 0.009 \\
                                        &                                             & 0.9007 \small± 0.020 & 0.8687 \small± 0.022 & 0.8147 \small± 0.027 \\
\cmidrule{2-5}
                                        & \multirow{2}{*}{Insurance}                  & 0.7963 \small± 0.014 & 0.7896 \small± 0.009 & 0.7089 \small± 0.002 \\
                                        &                                             & 0.8191 \small± 0.011 & 0.8000 \small± 0.004 & 0.6891 \small± 0.008 \\
\cmidrule{1-5}
\multirow{4}{*}{\makecell{Information \\ Technology}} & \multirow{2}{*}{Communication Equipment}    & 0.8598 \small± 0.016 & 0.8239 \small± 0.011 & 0.7653 \small± 0.016 \\
                                        &                                             & 0.8437 \small± 0.022 & 0.8171 \small± 0.011 & 0.7096 \small± 0.001 \\
\cmidrule{2-5}
                                        & \multirow{2}{*}{Semiconductor Equipment}    & 0.8243 \small± 0.015 & 0.8145 \small± 0.015 & 0.7505 \small± 0.010 \\
                                        &                                             & 0.8457 \small± 0.012 & 0.8019 \small± 0.016 & 0.7174 \small± 0.004 \\
\cmidrule{1-5}
\multirow{4}{*}{\makecell[l]{Communication \\ Services}} & \multirow{2}{*}{Telecommunications}         & 0.8826 \small± 0.006 & 0.8224 \small± 0.007 & 0.7338 \small± 0.004 \\
                                        &                                             & 0.8897 \small± 0.009 & 0.8656 \small± 0.010 & 0.7560 \small± 0.005 \\
\cmidrule{2-5}
                                        & \multirow{2}{*}{Media}                      & 0.8537 \small± 0.010 & 0.8348 \small± 0.010 & 0.7299 \small± 0.004 \\
                                        &                                             & 0.8751 \small± 0.011 & 0.8322 \small± 0.010 & 0.7402 \small± 0.008 \\
\cmidrule{1-5}
\multirow{4}{*}{Utilities}              & \multirow{2}{*}{Electric Utilities}         & 0.8941 \small± 0.009 & 0.8781 \small± 0.06 & 0.7885 \small± 0.002 \\
                                        &                                             & 0.8656 \small± 0.013 & 0.8346 \small± 0.015 & 0.7750 \small± 0.006 \\
\cmidrule{2-5}
                                        & \multirow{2}{*}{Gas Utilities}              & 0.8338 \small± 0.012 & 0.7953 \small± 0.011 & 0.7287 \small± 0.006 \\
                                        &                                             & 0.8052 \small± 0.013 & 0.7940 \small± 0.012 & 0.6800 \small± 0.008 \\
\cmidrule{1-5}
\multirow{4}{*}{Real Estate}            & \multirow{2}{*}{Equity Real Estate Trusts}  & 0.8208 \small± 0.015 & 0.7890 \small± 0.015 & 0.6868 \small± 0.006 \\
                                        &                                             & 0.8236 \small± 0.014 & 0.8163 \small± 0.005 & 0.7233 \small± 0.003 \\
\cmidrule{2-5}
                                        & \multirow{2}{*}{Management and Development} & 0.8277 \small± 0.021 & 0.8134 \small± 0.015 & 0.7358 \small± 0.015 \\
                                        &                                             & 0.8155 \small± 0.013 & 0.8052 \small± 0.013 & 0.6986 \small± 0.002 \\
\bottomrule
\end{tabular}
\end{table*}

\subsection{Detailed Diversity Analysis}

\begin{figure}[H]
    \centering
    \includegraphics[width=0.45\textwidth]{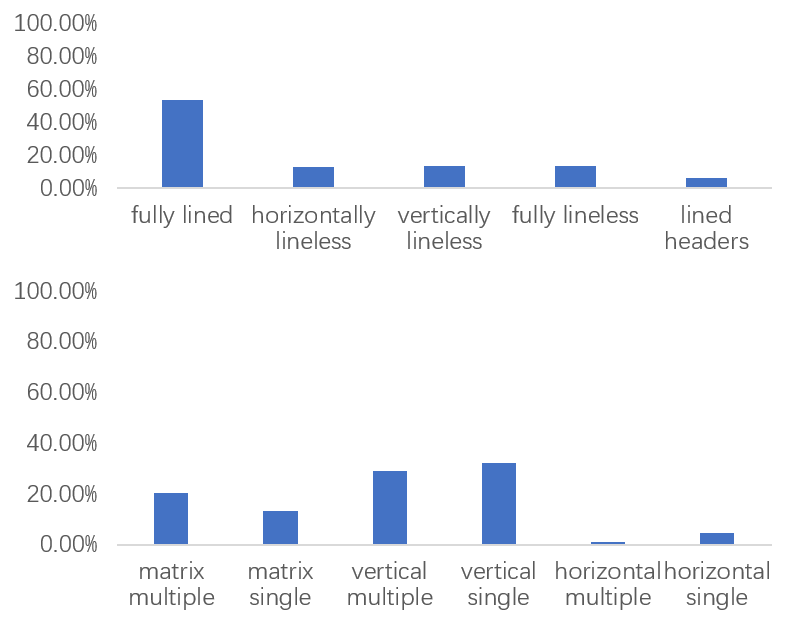}
    \caption{Detailed TableNet composition}
    \label{fig:detailed-composition}
\end{figure}

Based on the obeservation of crawled real-world tables, we divided tables into 5 types based on line type and 6 types based on table headers. As shown in Figure \ref{fig:detailed-composition}, our system is capable of generating given types of tables. In Figure \ref{fig:detailed-composition}, fully lined means every cell has complete borderlines, and horizontally/vertically lineless mean cells miss their horizontal/vertical border lines. Lined Headers means only header cells has borderlines. Matrix/verticle/horizontal tables means they have both/top headers/left headers.

However, existing datasets always fail to cover these types. For example, PubTabNet and FinTabNet is collected from scientific articles and is mostly consisted of verticle single header tables that only contain horizontal lines with black and white color. TableBank is mostly consisted of fully lined verticle single header tables which harms the ability of models to recognize lineless tables.

\section{Document Crawling Keywords}
\label{ap:kwd}

We crawled Microsoft search engine using Selenium by querying in format: 

\begin{center}
\texttt{company abbreviation + telecom-related keyword + filetype:pdf/doc}
\end{center}

Keywords are shown in Table \ref{tab:keywords}.

\begin{table*}
\caption{Keywords}
\label{tab:keywords}
\centering
\begin{tabular}{lll}
\toprule
\multicolumn{3}{l}{\textbf{Company Abbreviation}} \\
\midrule
\texttt{China Unicom} & \texttt{China Telecom} & \texttt{China Mobile} \\ 
\multicolumn{3}{l}{\texttt{China Broadnet}} \\
\midrule
\multicolumn{3}{l}{\textbf{Telecom-Related Keywords}} \\
\midrule
\texttt{5G} & \texttt{optical fiber} & \texttt{telecommunications} \\
\texttt{operator} & \texttt{base station} & \texttt{backbone network} \\
\texttt{gigabit} & \texttt{data center} & \texttt{FTTR} \\
\texttt{quality of service} & \texttt{roaming} & \texttt{value-added services} \\
\texttt{service hall} & \texttt{network access} & \texttt{voice services} \\
\texttt{broadband} & \texttt{cloud computing} & \texttt{local area network} \\
\texttt{Internet of Things} & \texttt{gateway services} & \texttt{ring back tone} \\
\texttt{network security} & \texttt{Wi-Fi} & \texttt{core network} \\
\texttt{network coverage} & \texttt{access network} & \texttt{network optimization} \\
\texttt{customer relationship management} & \texttt{IPTV} & \texttt{5G plans} \\
\texttt{data pricing} & \texttt{VoLTE} & \texttt{plan pricing} \\
\bottomrule
\end{tabular}
\end{table*}

\section{LLM HTML Generation Prompts}
\label{ap:html-prompts}                                     

\subsection{Structure-Only Generation Prompt}
\label{subap:structure-only-prompt}

Please generate an HTML table with \{no\_of\_rows\} rows and \{no\_of\_cols\} columns 
containing \{content\}. 
Only the \texttt{\textless tr\textgreater}, \texttt{\textless td\textgreater}, \texttt{\textless /tr\textgreater}, and \texttt{\textless /td\textgreater} tags should be used in the sequence.

Please return the generated table in the following JSON format: 
\{"html": "HTML table string"\}

The table should include cells that span multiple rows or columns.
Below are two matrices indicating the row spans and column spans:
Row span matrix: \{row\_spans\_matrix\}  
Column span matrix: \{col\_span\_matrix\}

\subsection{Topic Generation Prompt}

Please give me \{copy\} highly domain-relevant \{lang\} phrases related to the field of \{domain\}.  
Each phrase should be as detailed and specific as possible.  
Then return the generated phrases in the following JSON format:  
\{"phrase": [phrase1, phrase2, phrase3, phrase4, phrase5]\}
The following topics have already been used.  
Please do not repeat these topics or generate content that is highly similar to them:  
[Used topics]

\subsection{HTML Header Fulfilling Prompt}

Below is an HTML table in the domain of \{domain\}, with the topic: \{topic\}, and the table language is \{lang\}.  
The HTML sequence to be filled is as follows:  
\{HTML\_CODE\}

The structure of the HTML table is complete -- the header section (\textless th\textgreater\ tags) already contains the column names, while the body section (\textless td\textgreater\ tags) is empty.  
Without making any changes to the original HTML structure, attributes, or tags, please fill in the content of the table body (\textless td\textgreater\ tags) based on the given topic and existing headers. Ensure that each cell in the table contains distinct and meaningful content.

Generate \{copy\} different versions of the filled HTML table.  
Then return the results in the following JSON format:  
\texttt{\{"html": [list of filled HTML tables]\}}

\subsection{LLM Experiment and Few-Shot Prompt}
\label{subap:full-html-prompt}

Read the table content from the image and convert it into an HTML table format.  
The HTML must start with \textless html\textgreater\textless body\textgreater\textless table\textgreater\ and end with \textless /table\textgreater\textless /body\textgreater\textless /html\textgreater\, and the entire HTML should be output in a single line. Your output must contain only the HTML that begins with \textless html\textgreater\textless body\textgreater\textless table\textgreater\ and ends with \textless /table\textgreater\textless /body\textgreater\textless /html\textgreater\, and nothing else.

Examples:  
HTML1  
HTML2  
...

\subsection{HTML Body Fulfilling Prompt}

You will act as a strict HTML editor.  
You are tasked with processing a table in the \{domain\} domain, with the topic: \{topic\}, and the table language is \{lang\}.  
The HTML snippet to be completed is as follows:  
\{HTML\_CODE\}

The above is the original HTML paragraph for the table header.  
Without making any changes to the original HTML structure, please fill in the missing column names inside each \textless th\textgreater\ tag.

Important rules:

  Do not add, delete, or adjust any tags, attributes, or structure;
  Only fill in the content of \textless th\textgreater\ tags -- even if a \textless th\textgreater\ tag has colspan or rowspan attributes, you must still assign it an appropriate column name and must not leave it empty;
  The return format must be the following JSON structure:

\subsection{HTML Ranking Prompt}

You are an expert in table quality assessment, specializing in analyzing the structure, semantics, and topic relevance of HTML tables. Based on the input below, please evaluate the table’s quality across multiple dimensions and provide specific reasons for the scores.

Dimensions (1 to 5, integers only):

Structure Correctness (structure\_rank): (The score must be strictly \{score\}) \\
This score is based only on structural indicators; cell content does not affect this score. \\
The table's logical column count per row is detailed as follows: \{structure\_info\_prompt\}.

Topic Relevance (topic\_rank): \\
It is based only on the relevance between table content and the topic, ignoring expression or semantics. \\
Assess whether the content matches the topic "\{topic\}". \\
Ensure full coverage of the dimensions mentioned in the title (e.g., time, location, metrics, technologies, organizations), no more, no less. \\
Check whether the table reflects the following entity information: \{entities\_prompt\}. Inclusion of these entities—whether specific or general—is required. Irrelevant or completely unrelated content is prohibited. No fabricated or far-fetched justifications are allowed. \\
Do not rely solely on the literal presence of keywords; semantic alignment with the topic is sufficient.

Semantic Consistency (semantic\_rank): \\
Are there completely empty cells in the HTML? If so, suggest filling them with content consistent with the table headers. \\
Are the table headers and body semantically aligned? \{complex\_string2\} \\
Are the headers clear and detailed? \\
Are there garbled or corrupted characters? \\
Deduct points based on the proportion of cells exhibiting the above issues. Do not assign extremely low scores for isolated problems. \\
Content like "N/A", "-", or "TBD" is not considered empty.

Overall Score (rank): \\
This should be the lowest score among the three dimensions above.

Output Format should strictly be JSON format, no extra text. \\
Topic: \{topic\} \\
Raw HTML Table Code: \{html\_code\}

\subsection{Modified CoSyn Prompt}

You are a content creation expert with broad domain knowledge. \\
Please generate content material for me based on the following setup: \\

Topic: \{TOPIC\} \\
Figure Type: Table \\

Please follow the requirements below: \\

The generated material should be closely related to the topic and tailored according to the persona. The content structure should be suitable for generating the specified figure type (e.g., tables, charts, etc.). \\

All content must be factual and credible, using real-world entity names. The use of placeholders (such as xxA, xxB, {[}Name{]}, {[}Date{]}, etc.) is strictly prohibited. \\

The material should be diverse and cover the topic from different perspectives to ensure informational richness and breadth. \\

Control the amount of content and only provide key information so that it fits on a single-page document. \\

All content must be written in English, even if the persona is non-English-speaking. \\

Please output in JSON format, do not include any additional explanatory text.

You are an expert in writing HTML documents. \\
I have a dataset about \{TOPIC\} that can be used to generate an HTML table. \\
The data is as follows (in JSON format): \\
\textless data\textgreater \\
\{DATA\} \\
\textless /data\textgreater \\

Please use HTML and CSS to generate an HTML table. Follow the requirements below: \\

\textbf{Styling Requirements:} \\
(1) You may use any CSS frameworks, libraries, or tools to build the page. \\
(2) Be creative and ensure the webpage is distinctive in terms of typography, colors, borders, layout, etc., while aligning with the topic and target figure type. \\
(3) Use appropriate design proportions (e.g., margins, page size, content density, etc.) to ensure the information is clearly presented, with no overlapping text or layout issues. \\
(4) All content must be displayed within a single page---do not make the page too long or too sparse. This is a very important requirement. \\
(6) All text must be inside the table. \\
\textbf{[(7) Generated Table should contain rowspan or colspan attribute.]} \\

\textbf{Code Requirements:} \\
(1) Please hardcode the provided data directly into the HTML page. Do not use any backend calls. Ensure correct HTML syntax and formatting. \\
(2) Include both HTML and CSS in a single HTML file. Do not use any external resource files. \\

\textbf{Output Requirements:} \\
Please mark the code block. \\
Do not output any additional explanatory text---the output should be a single HTML file wrapped in \textless html\textgreater\textless /html\textgreater.

\newpage

\makeatletter
\@ifundefined{isChecklistMainFile}{
  % We are compiling a standalone document
  \newif\ifreproStandalone
  \reproStandalonetrue
}{
  % We are being \input into the main paper
  \newif\ifreproStandalone
  \reproStandalonefalse
}
\makeatother

\ifreproStandalone
\documentclass[letterpaper]{article}
\usepackage[submission]{aaai2026}
\setlength{\pdfpagewidth}{8.5in}
\setlength{\pdfpageheight}{11in}
\usepackage{times}
\usepackage{helvet}
\usepackage{courier}
\usepackage{xcolor}
\frenchspacing

\begin{document}
\fi
\setlength{\leftmargini}{20pt}
\makeatletter\def\@listi{\leftmargin\leftmargini \topsep .5em \parsep .5em \itemsep .5em}
\def\@listii{\leftmargin\leftmarginii \labelwidth\leftmarginii \advance\labelwidth-\labelsep \topsep .4em \parsep .4em \itemsep .4em}
\def\@listiii{\leftmargin\leftmarginiii \labelwidth\leftmarginiii \advance\labelwidth-\labelsep \topsep .4em \parsep .4em \itemsep .4em}\makeatother

\setcounter{secnumdepth}{0}
\renewcommand\thesubsection{\arabic{subsection}}
\renewcommand\labelenumi{\thesubsection.\arabic{enumi}}

\newcounter{checksubsection}
\newcounter{checkitem}[checksubsection]

\newcommand{\checksubsection}[1]{%
  \refstepcounter{checksubsection}%
  \paragraph{\arabic{checksubsection}. #1}%
  \setcounter{checkitem}{0}%
}

\newcommand{\checkitem}{%
  \refstepcounter{checkitem}%
  \item[\arabic{checksubsection}.\arabic{checkitem}.]%
}
\newcommand{\question}[2]{\normalcolor\checkitem #1 #2 \color{blue}}
\newcommand{\ifyespoints}[1]{\makebox[0pt][l]{\hspace{-15pt}\normalcolor #1}}

\section*{Reproducibility Checklist}

\vspace{1em}
\hrule
\vspace{1em}

\textbf{Instructions for Authors:}

This document outlines key aspects for assessing reproducibility. Please provide your input by editing this \texttt{.tex} file directly.

For each question (that applies), replace the ``Type your response here'' text with your answer.

\vspace{1em}
\noindent
\textbf{Example:} If a question appears as
\begin{center}
\noindent
\begin{minipage}{.9\linewidth}
\ttfamily\raggedright
\string\question \{Proofs of all novel claims are included\} \{(yes/partial/no)\} \\
Type your response here
\end{minipage}
\end{center}
you would change it to:
\begin{center}
\noindent
\begin{minipage}{.9\linewidth}
\ttfamily\raggedright
\string\question \{Proofs of all novel claims are included\} \{(yes/partial/no)\} \\
yes
\end{minipage}
\end{center}
Please make sure to:
\begin{itemize}\setlength{\itemsep}{.1em}
\item Replace ONLY the ``Type your response here'' text and nothing else.
\item Use one of the options listed for that question (e.g., \textbf{yes}, \textbf{no}, \textbf{partial}, or \textbf{NA}).
\item \textbf{Not} modify any other part of the \texttt{\string\question} command or any other lines in this document.\\
\end{itemize}

You can \texttt{\string\input} this .tex file right before \texttt{\string\end\{document\}} of your main file or compile it as a stand-alone document. Check the instructions on your conference's website to see if you will be asked to provide this checklist with your paper or separately.

\vspace{1em}
\hrule
\vspace{1em}

% The questions start here

\checksubsection{General Paper Structure}
\begin{itemize}

\question{Includes a conceptual outline and/or pseudocode description of AI methods introduced}{(yes/partial/no/NA)}
yes

\question{Clearly delineates statements that are opinions, hypothesis, and speculation from objective facts and results}{(yes/no)}
yes

\question{Provides well-marked pedagogical references for less-familiar readers to gain background necessary to replicate the paper}{(yes/no)}
yes

\end{itemize}
\checksubsection{Theoretical Contributions}
\begin{itemize}

\question{Does this paper make theoretical contributions?}{(yes/no)}
no

	\ifyespoints{\vspace{1.2em}If yes, please address the following points:}
        \begin{itemize}
	
	\question{All assumptions and restrictions are stated clearly and formally}{(yes/partial/no)}
	Type your response here

	\question{All novel claims are stated formally (e.g., in theorem statements)}{(yes/partial/no)}
	Type your response here

	\question{Proofs of all novel claims are included}{(yes/partial/no)}
	Type your response here

	\question{Proof sketches or intuitions are given for complex and/or novel results}{(yes/partial/no)}
	Type your response here

	\question{Appropriate citations to theoretical tools used are given}{(yes/partial/no)}
	Type your response here

	\question{All theoretical claims are demonstrated empirically to hold}{(yes/partial/no/NA)}
	Type your response here

	\question{All experimental code used to eliminate or disprove claims is included}{(yes/no/NA)}
	Type your response here
	
	\end{itemize}
\end{itemize}

\checksubsection{Dataset Usage}
\begin{itemize}

\question{Does this paper rely on one or more datasets?}{(yes/no)}
yes

\ifyespoints{If yes, please address the following points:}
\begin{itemize}

	\question{A motivation is given for why the experiments are conducted on the selected datasets}{(yes/partial/no/NA)}
	yes

	\question{All novel datasets introduced in this paper are included in a data appendix}{(yes/partial/no/NA)}
	yes

	\question{All novel datasets introduced in this paper will be made publicly available upon publication of the paper with a license that allows free usage for research purposes}{(yes/partial/no/NA)}
	yes

	\question{All datasets drawn from the existing literature (potentially including authors' own previously published work) are accompanied by appropriate citations}{(yes/no/NA)}
	yes

	\question{All datasets drawn from the existing literature (potentially including authors' own previously published work) are publicly available}{(yes/partial/no/NA)}
	yes

	\question{All datasets that are not publicly available are described in detail, with explanation why publicly available alternatives are not scientifically satisficing}{(yes/partial/no/NA)}
	NA

\end{itemize}
\end{itemize}

\checksubsection{Computational Experiments}
\begin{itemize}

\question{Does this paper include computational experiments?}{(yes/no)}
yes

\ifyespoints{If yes, please address the following points:}
\begin{itemize}

	\question{This paper states the number and range of values tried per (hyper-) parameter during development of the paper, along with the criterion used for selecting the final parameter setting}{(yes/partial/no/NA)}
	no

	\question{Any code required for pre-processing data is included in the appendix}{(yes/partial/no)}
	yes

	\question{All source code required for conducting and analyzing the experiments is included in a code appendix}{(yes/partial/no)}
	yes

	\question{All source code required for conducting and analyzing the experiments will be made publicly available upon publication of the paper with a license that allows free usage for research purposes}{(yes/partial/no)}
	yes
        
	\question{All source code implementing new methods have comments detailing the implementation, with references to the paper where each step comes from}{(yes/partial/no)}
	yes

	\question{If an algorithm depends on randomness, then the method used for setting seeds is described in a way sufficient to allow replication of results}{(yes/partial/no/NA)}
	NA

	\question{This paper specifies the computing infrastructure used for running experiments (hardware and software), including GPU/CPU models; amount of memory; operating system; names and versions of relevant software libraries and frameworks}{(yes/partial/no)}
	partial

	\question{This paper formally describes evaluation metrics used and explains the motivation for choosing these metrics}{(yes/partial/no)}
	yes

	\question{This paper states the number of algorithm runs used to compute each reported result}{(yes/no)}
	yes

	\question{Analysis of experiments goes beyond single-dimensional summaries of performance (e.g., average; median) to include measures of variation, confidence, or other distributional information}{(yes/no)}
	yes

	\question{The significance of any improvement or decrease in performance is judged using appropriate statistical tests (e.g., Wilcoxon signed-rank)}{(yes/partial/no)}
	yes

	\question{This paper lists all final (hyper-)parameters used for each model/algorithm in the paper’s experiments}{(yes/partial/no/NA)}
	yes

\end{itemize}
\end{itemize}
\ifreproStandalone
\end{document}
\fi
\end{document}